\documentclass[twocolumn,showpacs,preprintnumbers,amsmath,amssymb]{revtex4}

\usepackage{graphicx}
\usepackage{dcolumn}
\usepackage{bm}

\begin{document}

\title{Imaging in scattering media via the second-order correlation of light field}

\author{Wenlin Gong}
\author{Pengli Zhang}
\author{Xia Shen}
\author{ Shensheng Han}
\email{sshan@mail.shcnc.ac.cn} \affiliation{ Key Laboratory for
Quantum Optics and Center for Cold Atom Physics of CAS, Shanghai
Institute of Optics and Fine Mechanics, Chinese Academy of Sciences,
Shanghai 201800, China }

\date{\today}

\begin{abstract}
Imaging with the second-order correlation of two light fields is a
method to image an object by two-photon interference
involving a joint detection of two photons at distant space-time
points. We demonstrate for the first time that an image with high
quality can still be obtained in the scattering media by applying
the second-order correlation of illuminating light field. The
scattering effect on the visibility of images is analyzed both
theoretically and experimentally. Potential applications and the
methods to further improve the visibility of the images in
scattering media are also discussed.
\end{abstract}

\pacs{42.50.Ar, 42.68.Mj, 42.50.Dv, 42.62.Be}
\maketitle

Multiple scattering has a great influence on the quality of images.
The information will be degraded and the images suffer reduced
resolution and contrast because of multiple scattering, such as
light propagation and imaging in the atmosphere \cite{Mckechnie},
neutron imaging \cite{Raine}, imaging and diagnosis in life and
medical sciences \cite{Maher}. In medical and clinic diagnosis,
comparing with X-ray, optical photons provide nonionizing and safe
radiation for medical applications, and are now becoming an
increasing interesting method for imaging in biological tissues.
Such as Optical coherence tomography (OCT), Diffuse optical
tomography (DOT), Photoacoustic tomography (PAT)and so on
\cite{Wang,Gibson,Yodh,Arridge,Zhang,Hsiung}. Although the quality
of images in scattering media have an enhancement in some extent by
these techniques, there are still lots of problems which are
difficult to be settled. Because all conventional imaging methods
are based on first-order correlation of light field, we ``see" an
image only when we look at the object, which means the detection and
imaging are unseparated in conventional imaging process. So when the
information of the object is distorted by multiple scattering, and
the information of both multiple scattering and the object are
unknown, we can not, in principle, obtain exactly an image destroyed
by multiple scattering.

With respect to the classical area of imaging, the field of quantum
imaging aims to devise novel techniques for optical imaging, by
exploiting the quantum nature of light \cite{Gatti}. Based on the
characteristic of Bose-Einstein correlation of light fields and the
theory of two-photon interference \cite{Glauber}, the imaging method
by the second-order correlation of two light fields, for example,
ghost imaging with entangled source or thermal light, becomes a new
kind of imaging technique \cite{Angelo}. In ghost imaging schemes,
even if the test detector is a pointlike or bucket
detector, by measuring the second-order correlation function of the
two light fields, we can still obtain the image of the object with
both entangled source and thermal light by scanning the position of
the photons which never actually passed through the object
\cite{Pittman,Valencia,Cheng,Gatti1,Gatti2,Bennink2,Zhang1,Zhang2}.
This imaging method, for the first time in the history of imaging
technology development, leads to the separation of detection and
imaging. In this letter, when there is multiple scattering in the test path but no multiple
scattering in the reference path, imaging in the scattering
media with thermal light is investigated by the second-order
correlation of two light fields.

Fig. 1 shows the schematic of experimental setup. The thermal light
source $S$ produced by the method described in Ref.
\cite{Zhang2,Martienssen,Liu}, first propagates through a beam
splitter, then is divided into a test and a reference path. In the
test path, the light goes through a thin lens of focal length $f_1$,
scattering media and then to the test detector $D_t$. In the reference
path, the light propagates through another thin lens of focal length
$f$ then to the CCD camera $D_r$ with the axial resolution 6.45$\times$6.45$\mu$m.

\begin{figure}
\centerline{
\includegraphics[width=8.5cm]{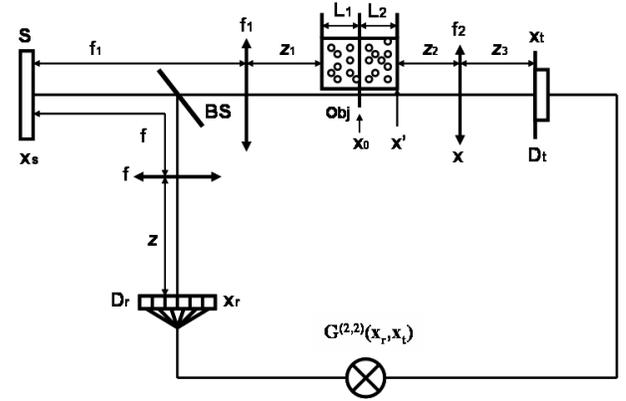}}
\caption{Scheme for the second-order correlation with thermal light in
the scattering media.}
\end{figure}

In Fig. 1, The impulse response function between the plane $x$ and
the plane $x_0$ is \cite{Wang,Vesperinas,Holstein}
\begin{eqnarray}
h(x,x_0 ) = \alpha h_{in} (x,x_0 ) + \beta h_{sca} (x,x_0 ).
\end{eqnarray}
\begin{eqnarray}
\left| \alpha  \right|^2  + \left| \beta  \right|^2  = 1.
\end{eqnarray}
where $h_{in}(x,x_0)$ is the impulse response function with no
scattering media, and $h_{sca}(x,x_0)$ is the impulse response
function because of the interactions of multiple scattering.
$\alpha$, $\beta$ are the probability amplitudes of incident light
and scattering light, respectively.

The probability distribution function in scattering media can be
represented by point scattering function(PSF). Generally, there are
two forms of PSF: Lorentzian-shaped and Gaussian-shaped distribution
\cite{Hassanein,Segre}.

\begin{subequations}
\label{eq:whole}
\begin{eqnarray}
h_{sca} (x,x_0 ) \propto \int dx' P(x',x_0 )_{L_A}
h(x,x')_{(L_{2}+z_{2})},
\end{eqnarray}
\begin{eqnarray}
P(x',x_0 )_{L_A} =[\frac{2}{{\pi \Delta x_{L_A } ^2 }}]^{1/4} \exp
\left\{ { - (\frac{{x' - x_0 }}{{\Delta x_{L_A} }})^2 } \right\},
\end{eqnarray}
\begin{eqnarray}
\int {\left| {P(x',x_0 )_{L_A} } \right|} ^2 dx'=1.
\end{eqnarray}
\begin{eqnarray}
\beta  \propto \frac{{D^{a_\beta  } w^{c_\beta  } {L_A}^{d_\beta  } n^{e_\beta  } }}{{\lambda ^{b_\beta  } }};\Delta x_{L_A }  \propto \frac{{D^{a_x } w^{c_x } {L_A}^{d_x } n^{e_x } }}{{\lambda ^{b_x } }}.
\end{eqnarray}
\end{subequations}
here, $P(x',x_0 )_{L_A}$ is the Gaussian-shaped point scattering
probability amplitude from the position $x_0$ to $x'$ when the light
goes through the scattering medium with effective thickness $L_A$.
$\Delta x_{L_A}$, $D$, $w$, $\lambda$ and $n$ are the broadening width,
the diameter size of scattering particles, the concentration of suspended particles,
the wavelength of the incident light and refractive index of the scattering medium, respectively.
$\alpha$, $\beta$ and $\Delta x_{L_A}$ are all determined
by specific experimental conditions.

By optical coherence theory \cite{Cheng,Gatti1,Glauber}, the
second-order correlation function between the two detectors is:

\begin{eqnarray}
\Delta G^{(2,2)} (x_r ,x_t ) =  < I(x_r )I(x_t ) >  -  < I(x_r ) > <
I(x_t ) >\nonumber\\ =\left| \right.\int {dx_1 } \int {dx_2}
G^{(1,1)} (x_1 ,x_2 ) h_r ^ *  (x_1 ,x_r )h_t (x_2 ,x_t )\left.
\right|^2 .
\end{eqnarray}
where $< >$ denotes statistical average of the ensemble, and
$G^{(1,1)} (x_1 ,x_2)$ is the first-order correlation function on
the source plane, $h_t(x_t ,x_2)$ is the impulse response function
in the test path whereas $h_r^*(x_r,x_1)$ denotes phase conjugate of
the impulse response function in the reference path.

Suppose the light source is fully spatially incoherent,
\begin{equation}
G^{(1,1)} (x_1 ,x_2 ) = I_0 \delta (x_1  - x_2 ).
\end{equation}
where $I_0$ is a constant, and $\delta(x)$ is Dirac delta function.

Under the paraxial approximation, and when the effective apertures
of the lenses in the optical system are large enough, the impulse
response function of the reference system is
\begin{equation}
h_r (x_r ,x_1) \propto \exp \left\{ {\frac{{j\pi }}{{\lambda f}}(1 -
\frac{z}{f})x_1 ^2  - \frac{{2j\pi }}{{\lambda f}}x_r x_1 }
\right\}.
\end{equation}
when the object plane, the thin lens $f_2$ and the test detector plane satisfy the thin lens equation
\begin{equation}
\frac{1}{{z_2+L_2 }} + \frac{1}{{z_3 }} = \frac{1}{{f_2 }}.
\end{equation}
then the impulse response function in the test path is
\begin{subequations}
\label{eq:whole}
\begin{eqnarray}
h_t(x_t ,x_2 ) \propto \int {dx'} [\alpha _1 \exp \left\{ { -
\frac{{2j\pi }}{{\lambda f_1 }}x'x_2 } \right\} + \beta _1 \int d
x_2 '\nonumber\\\times P(x',x_2 ')_{L_{1A} } \exp \left\{ {-
\frac{{2j\pi }}{{\lambda f_1 }}x_2 'x_2 } \right\}]
t(x')C(x')\nonumber\\\times\exp \left\{ {\frac{{j\pi }}{{\lambda f_1
}}(1 - \frac{{z_1 + L_1 }}{{f_1 }})x_2 ^2 } \right\}.
\end{eqnarray}
\begin{eqnarray}
C(x') = \alpha _2 \delta (x' + \frac{{z_2  + L_2 }}{{z_3 }}x_t ) + \beta _2 P( - \frac{{z_2  + L_2 }}{{z_3 }}x_t ,x')_{L_{2A} }.
\end{eqnarray}
\end{subequations}
where t(x) is the transmission function of the object. If
\begin{eqnarray}
\frac{{1 - \frac{z}{f}}}{f} = \frac{{1 - \frac{{z_1  + L_1 }}{{f_1
}}}}{{f_1 }}.
\end{eqnarray}
and $\frac{{f_1 }}{f} = \frac{{z_2  + L_2 }}{{z_3 }}$, then the same speckle distributions
can be obtained on the two detector planes without the scattering media and the object.
Substituting Eqs. (5)-(9) into Eq. (4), the correlation function is
\begin{eqnarray}
 \Delta G^{(2,2)} (x_r ,x_t ) \propto \left| {\{ \alpha _1 \alpha _2 \delta (x_r  + x_t ) + P( - \frac{{f_1 }}{f}x_t ,\frac{{f_1 }}{f}x_r )_{L_{2A} } } \right.\nonumber\\
  \times \beta _1 \alpha _2 \} t( - \frac{{f_1 }}{f}x_t ) + \alpha _1 \beta _2 P( - \frac{{f_1 }}{f}x_t ,\frac{{f_1 }}{f}x_r )_{L_{1A} } t(\frac{{f_1 }}{f}x_r )\nonumber\\
 \left. { + \beta _1 \beta _2 \int {dx'(x')} P(x',\frac{{f_1 }}{f}x_r )_{L_{1A} } P( - \frac{{f_1 }}{f}x_t ,x')_{L_{2A} } } \right|^2.
\end{eqnarray}
If the test detector is a bucket detector, integrating $x_t$ in Eq. (10), then
\begin{eqnarray}
 \Delta G^{(2)} (x_r ) = \int {\Delta G^{(2,2)} (x_r ,x_t )dx_t }.
\end{eqnarray}
If the test detector is a CCD camera, because the images on the two camera planes are inverse, in the case of $x_t=-x_r$, Eq. (10) can be rewritten as
\begin{eqnarray}
\Delta G^{(2,2)} (x_r , - x_r ) \propto \left| {\left\{ {\alpha _1
\alpha _2  + \alpha _1 \beta _2 \left[ {\frac{2}{{\pi \Delta
x_{L_{2A} } ^2 }}} \right]^{1/4} } \right.} \right.\nonumber\\\left.
{ + \beta _1 \alpha _2 \left. {\left[ {\frac{2}{{\pi \Delta
x_{L_{1A} } ^2 }}} \right]^{1/4} } \right\}t(\frac{{f_1 }}{f}x_r ) +
\beta _1 \beta _2 C_0 } \right|^2 .
\end{eqnarray}

\begin{eqnarray}
C_0 = \int {dx't(x')} P(x',\frac{{f_1 }}{f}x_r )_{L_{1A} } P(
\frac{{f_1 }}{f}x_r ,x')_{L_{2A} }.
\end{eqnarray}

From Eq. (12)-(13), for $L_1$=0 (namely, $\beta_1$=0) or $L_2$=0 (namely, $\beta_2$=0), images with high quality can always be reconstructed by the second-order correlation of two light fields in scattering media. However, if the
object is fixed in the middle of the scattering media, the quality of the image is the worst.

In the experiments, we prepare a suspension liquid composed by
emulsion polymerization particles with particle diameter
$D$=3.26$\mu$m and the solution $NaCl$ with density $\rho$=1.19
$g/cm^3$. The vessel used to put the suspension liquid is designed
as 40mm$\times$10mm$\times$20mm. We take $\lambda$=650nm,
$f_1$=400.0mm, $f$=150.0mm, $f_2$=250.0mm, $z$=211.0mm,
$z_1$=300.0mm, $z_2+L_2$=390.0mm, $z_3$=243.8mm. The average sizes of
the speckles on the CCD camera plane are $40.6\mu$m. When the light goes
through the scattering media with thickness $L_1+L_2$=40mm,
$|\frac{\beta}{\alpha}|^2$$\approx$694, $\Delta
x_{L_A}$$\approx$1.36mm, the scattering coefficient of the medium
$\mu_s$$\approx$1.64/cm. The minimum characteristic scale of the
object (``\textbf{zhong}" ring) is 60$\mu$m and the diameter of the
ring is 1.6mm.

\begin{figure}
\centerline{
\includegraphics[width=8.5cm]{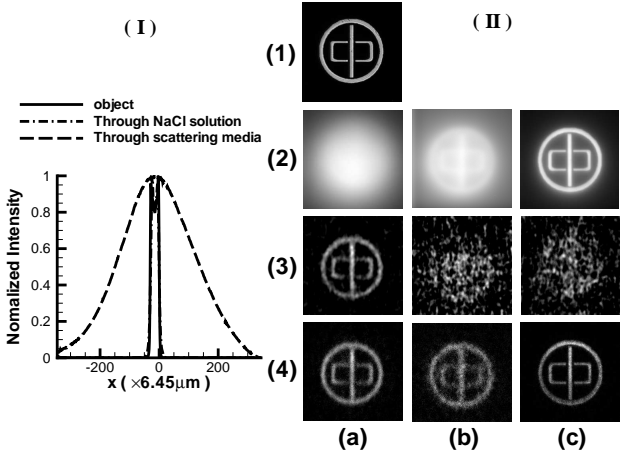}}
\caption{(I). The axial sections of the images when the thermal light
from a single slit (the silt width a=0.2mm) fixed before the mediums
went through different mediums with the scattering thickness $L_1+L_2$=40mm.
(II). Images of the object obtained when the sample was fixed in different
positions of the scattering media (averaged 15, 000 speckle frames). (a). $L_1$=0mm, $L_2$=40mm, $\triangle x_{L_{A}}$=1.36mm; (b).
$L_1$=20mm, $L_2$=20mm, $\triangle x_{L_{A}}$=0.74mm; and (c). $L_1$=40mm, $L_2$=0mm, $\triangle x_{L_{A}}$=0.01mm. (1). the
object; (2). Conventional imaging; (3)-(4). were images reconstructed by the second-order
correlation between two paths when the test detector was a bucket detector,
and a CCD camera with the axial resolution 6.45$\times$6.45$\mu$m, respectively.}
\end{figure}

Fig. 2(I) represents the point scattering functions when the thermal light goes through different mediums.
The Gaussian envelop will become wider because of multiple scattering. For conventional imaging, as shown in Fig. 2(II)(2), the quality
of images will reduce as the increase of the scattering thickness
$L_2$. Oppositely, when the test detector is a bucket detector, the quality of the images
reconstructed by the second-order correlation between two paths will decay as the scattering thickness
$L_1$ is increased [Fig. 2(II)(3)]. However, as shown in Fig. 2(II)(4-c),
if the test detector is a CCD camera with high axial resolution,
we can also obtain images with high
quality via the second-order correlation of the two light fields when there is only strong multiple scattering
between the object plane and the source. In addition, if the
object is fixed in the middle of the scattering media, the
visibility of the image will reduce, but the resolution doesn't
decay [Fig. 2(II)(4-b)].

\begin{figure}
\centerline{
\includegraphics[width=8.5cm]{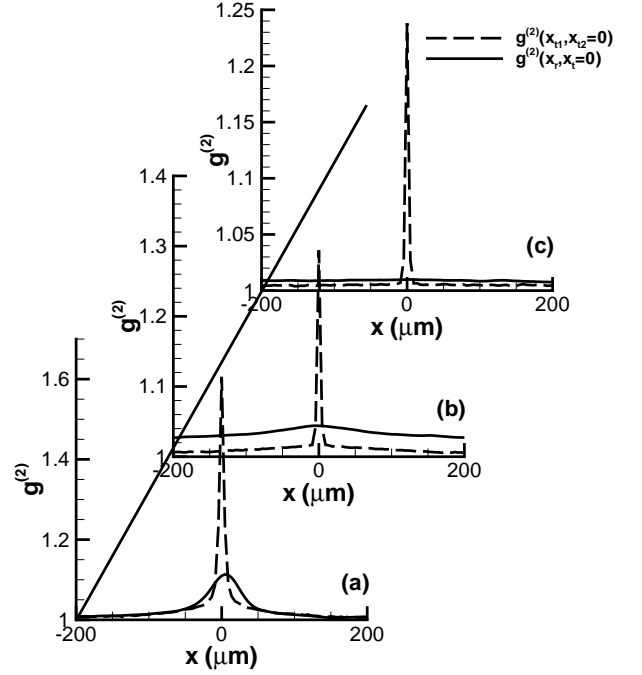}}
\caption{Effect of multiple scattering between the source and the
object plane on $g^{(2)}(x_{t1},x_{t2}=0)$ and
$g^{(2)}(x_r,x_t=0)$ for different scattering thickness $L_1$ and $L_2$=0 without the object.
(a). $L_1$=10mm; (b). $L_1$=20mm; (c). $L_1$=40mm.}
\end{figure}

The visibility of the images reconstructed by the second-order correlation of two light fields can be
explained by normalized second-order correlation function \cite{Glauber}
\begin{eqnarray}
g^{(2)} (x_1 ,x_2  = 0) = 1 + \frac{{\Delta G^{(2,2)} (x_1 ,x_2  =
0)}}{{ < I(x_1 ) >  < I(x_2  = 0) > }}.
\end{eqnarray}
where $g^{(2)} (x_1 ,x_2  = 0)$ denotes the second-order cross
correlation degree between two spatial positions $x_2$=0 and $x_1$.
For the thermal light field, $g^{(2)}(x_{t1},x_{t2}=0)$ reveals the
fluctuation of the light field, whereas $g^{(2)}(x_r,x_t=0)$
describes the cross correlation between two paths. As shown in Fig. 3, both
the maximum values of the second-order correlation degree
$g^{(2)}(x_{t1}=0,x_{t2}=0)$ and $g^{(2)}(x_r=0,x_t=0)$
decrease sharply as the scattering thickness $L_1$ is increased,
so the visibility of images will reduce (see Fig. 2(II)(3)). However,
even $g^{(2)}(x_r=0,x_t=0)$ is lower than 1.05, an image with
high quality can still be reconstructed when both
the detectors in two paths are CCD cameras with high axial resolution (see Fig. 2(II)(4)).

Because multiple scattering between the source and
the object plane destroys the cross-correlation between the two paths, thus
the visibility of the image and $g^{(2)}(x_r=0,x_t=0)$ will be
degraded as the increase of multiple scattering. Higher $g^{(2)}(x_r=0,x_t=0)$ and larger transverse coherent width on the object
plane can both enhance the visibility of the image in
scattering media reconstructed by the second-order correlation between the test and reference paths \cite{Shen,Gatti3}.

In addition, by the results described by Eq. (12) and represented in Figs. 2-3, when the both
detectors in two paths are CCD cameras with high axial resolution, the quality of images
obtained by the second-order correlation of two light fields is
also determined by the quality of images registered by the CCD camera $D_t$.
Thus, almost all existing conventional imaging schemes in scattering media can be applied in
the test path to further improve the quality of the correlation
imaging system. When $L_1<<L_2$, the quality of the image reconstructed by the second-order correlation of the two light fields is mainly determined by the mechanism of ``ghost" imaging. Because of the separation of detection and imaging for ``ghost" imaging, the test detector is only used to collect the radiation of the thermal light propagating through the object, while imaging system is located in the reference path. Because there is no multiple scattering in the reference path, the quality of the image isn't influenced. If $L_1>>L_2$, the quality of the reconstructed image mainly depends on the conventional imaging system in the test path. Because the scattering between the object and the CCD camera $D_t$ is week, thus the image with good quality can also be obtained.
Generally speaking, by the second-order correlation
of the two light fields, we can always obtain an image with much
better quality than the image achieved only with conventional
first-order correlation optical imaging methods in scattering media.
Entangled source and other nonclassical light source with higher
$g^{(2)}(x_{t1}=0,x_{t2}=0)$ may be applied to further improve
the quality of the image obtained by the second-order correlation
imaging system in scattering media \cite{Chan,Walls,Wang1}.

In conclusion, imaging via the second-order correlation of two light fields
provides a brand-new route for imaging in scattering media. We
demonstrate for the first time that we can always obtain an image
with much better quality reconstructed by the second-order correlation imaging
method than the image achieved by conventional first-order
correlation imaging methods in scattering media. This will be very
useful to imaging, test and diagnosis of biological tissues with
infrared and near infrared light.

The work was partly supported by the Hi-Tech Research and
Development Program of China under Grant Project No. 2006AA12Z115,
and Shanghai Fundamental Research Project under Grant Project No.
06JC14069.

\end{document}